\documentclass[aps,pra]{revtex4-1}
\usepackage[centertags]{amsmath}
\usepackage{amsfonts}
\usepackage{amssymb}
\usepackage{newlfont}
\usepackage{ae} 
\usepackage[T1]{fontenc}
\usepackage{graphicx}
\usepackage{subfigure}
\usepackage[ansinew]{inputenc}
\usepackage{lineno}
\usepackage{braket}
\usepackage{hyperref}

\newcommand{\Tr}{\operatorname{Tr}}
\newcommand{\abs}[1]{\left| #1 \right|} 
\newcommand{\matrixel}[3]{\left< #1 \vphantom{#2#3} \right|
 #2 \left| #3 \vphantom{#1#2} \right>} 

\begin{document}

\title{Quantum Interference between a Single-Photon Fock State and a Coherent State}
\newcommand{\atomicInst}{Vienna University of
Technology, Atomic Institute, Stadionallee 2, 1020 Vienna, Austria}
\newcommand{\AIT}{AIT Austrian Institute of Technology GmbH, Safety \& Security
Department, Optical Quantum Technologies, Donau-City-Str.~1, 1220
Vienna, Austria}

\author{A. Windhager} \affiliation{\atomicInst}

\author{M. Suda}
\email[corresponding author: ]{Martin.Suda@ait.ac.at}
\affiliation{\atomicInst}
\affiliation{\AIT}

\author{C. Pacher} \affiliation{\AIT}

\author{M. Peev} \affiliation{\AIT}

\author{A. Poppe} \affiliation{\AIT}

\begin{abstract}
We derive analytical expressions for the single mode quantum field
state at the individual output ports of a beam splitter when a
single-photon Fock state and a coherent state are incident on the
input ports. The output states turn out to be a statistical mixture
between a displaced Fock state and a coherent state. Consequently we
are able to find an analytical expression for the corresponding
Wigner function. Because of the generality of our calculations the
obtained results are valid for all passive and lossless optical four
port devices. We show further how the results can be adapted to the
case of the Mach-Zehnder interferometer. In addition we consider the
case for which the single-photon Fock state is replaced with a
general input state: a coherent input state displaces each general
quantum state at the output port of a beam splitter with the  displacement
parameter being the amplitude of the coherent state.
\\
\\
PACS. 42.50.Dv\\
Key words: coherent state, displaced Fock state, Mach-Zehnder,
entanglement, quantum interference
\end{abstract}
\maketitle

\section{Introduction}

Quantum optics is an exciting field, in which many fundamental
experiments, revealing the peculiarities of quantum mechanics, have
been conducted. The advantage of optical experiments, compared with
other fundamental experiments, is their simplicity. Quantum optics
has been one of the main vehicles in the development of quantum information technologies and in particular of
quantum cryptography
\cite{quantumcryptography,continuousvariableQKD,Chuang,Gisin} and
optical quantum computing
\cite{quantcomputing1,quantcomputing2,Chuang}.
The most widely used
quantum states in this respect are coherent states and Fock states
\cite{M&W}. Important basic building blocks in quantum optics are
beam splitters and Mach-Zehnder interferometers, which are passive
and lossless four port devices \cite{BS2,Yurke,Knight}. Many cases
of interference between vacuum, Fock states and coherent states have
already been theoretically studied \cite{Knight}. A famous example
is the so called Hong-Ou-Mandel effect \cite{HongOuMandel} where two
single-photon Fock states arrive simultaneously at each input port
of a balanced beam splitter. In this case there are no coincident
photons at the output ports. On the other hand interference between
a fluorescent photon and a classical field has been investigated
\cite{HOM88} where the photon is created from a coherently excited
atom going through its Rabi cycle of oscillation. This results in a
time dependence of the interferometric fringe visibility as a
function of the atomic Rabi frequency \cite{Bali93}.

Without going into details of the time-dependent photon generation a
somewhat simpler but nevertheless important problem arises when we
consider the interference between a true single-photon Fock
$|1\rangle$ state and a coherent state $|\alpha\rangle$. Based on an
experiment in 2002 a displaced Fock state has been synthesized to a
good approximation by overlapping a single-photon Fock state with a
strong coherent pulse on a highly reflective beam splitter
\cite{Lvovsky}. To the best of our knowledge an exact analytical
expression for the evolving quantum state in this experiment has not
been published so that the reconstructed state (Wigner function) of
the beam splitter output could be compared with the theory.


In this work we calculate analytically the
output of a beam splitter (BS) in the case of a single-photon Fock
state $\ket{1}$ and a coherent state $\ket{\alpha}$ impinging respectively on its two input
ports. In particular we derive an exact expression for the quantum
state of the separate beam splitter output ports. Because of the
generality of our calculations, the results we obtain are not
restricted to the beam splitter, but are valid for all passive and
lossless optical four port devices.  In addition we consider the case for which the
single-photon Fock state is replaced with a general input state. We
show that a coherent input state displaces the quantum state at the
output port of a beam splitter in phase space, whereby the displacement parameter is the
amplitude of the coherent state.
In the second part of this paper
we show how the results can be adapted to the case of a Mach-Zehnder
interferometer (MZI). Additionally, the mean photon numbers at the output
ports of the MZI are determined. These results serve as a first step
of investigating more complex systems that might find application in optical quantum computation.

The intention of this article is
to present the matter in a concise and didactic way.

\begin{figure}[htbp] 
   \centering
   {\includegraphics[width=0.34\textwidth]{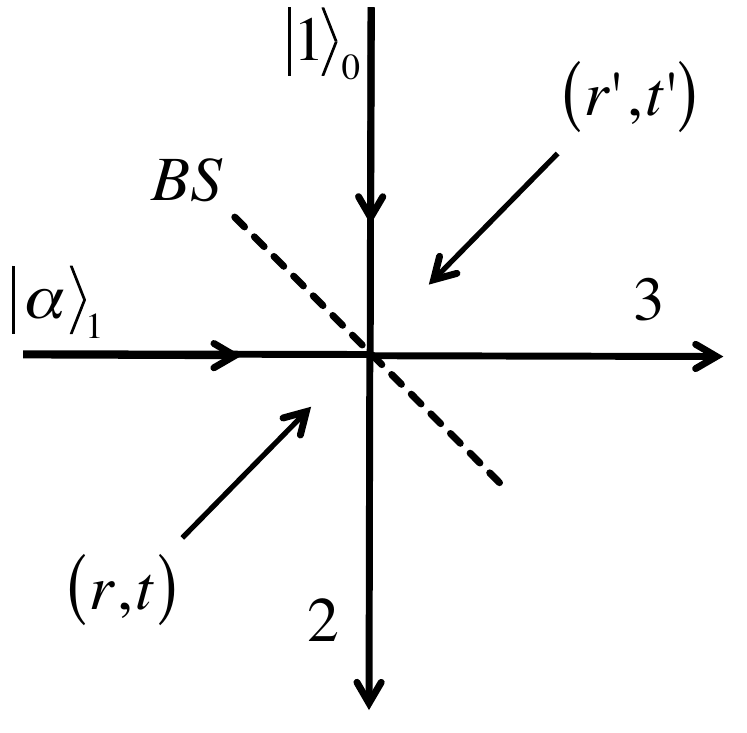}}
   \caption{Beam Splitter (BS) with a single-photon Fock state $\ket{1}_0$ and a
   coherent state $\ket{\alpha}_1$ incident on its input ports.}
   \label{fig:BS}
\end{figure}

\section{Interference at a beam splitter}

We consider a general beam splitter with complex reflection
coefficients $r$ and $r'$ and transmission coefficients $t$ and
$t'$. The phase relation between the coefficients depends on the
construction of the beam splitter \cite{BSHamilton}. In the
Heisenberg picture the annihilation operators $\hat a$ of the
incident fields transform as \cite{Knight}
\begin{eqnarray}
\left(
\begin{array}
[c]{cc}
\hat{a}_{2}\\
\hat{a}_{3}
\end{array}
\right)\,\,\,= \left(
\begin{array}
[c]{cc}
t' & r\\
r' & t
\end{array}
\right) \left(
\begin{array}
[c]{cc}
\hat{a}_{0}\\
\hat{a}_{1}
\end{array}
\right)= B \left(
\begin{array}
[c]{cc}
\hat{a}_{0}\\
\hat{a}_{1}
\end{array}
\right),\label{BSMatrix}
\end{eqnarray}
where the indices indicate the corresponding ports (modes). The
unitary scattering matrix $B$ must satisfy - for lossless devices -
the so called reciprocity relations due to Stokes \cite{Stokes}
\begin{align}\label{reci}
\abs{r'}=\abs{r}, \quad \abs{t}=\abs{t'}, \quad \abs{r}^2 +
\abs{t}^2 = 1 \,\,\,,\,\,\, r^*t'+r't^* = 0 .
\end{align}

\subsection{Input: Fock state and Coherent state}

In our setup (see Fig.~\ref{fig:BS}) the incident field states are a single-photon
Fock state $\ket{1}_0$ and a coherent state $\ket{\alpha}_1$ which can be written
as the following product state:
\begin{equation}\label{inputsingle}
\ket{1}_0 \ket{\alpha}_1 = \hat{D}_1(\alpha) \hat a_0^{\dagger}  \ket{0}_0 \ket{0}_1 =
e^{\alpha \hat a_1^{\dagger}-\alpha^*\hat a_1} \hat a_0^{\dagger}  \ket{0}_0 \ket{0}_1,
\end{equation}
where $\hat a^{\dagger}_0$ is the creation operator and  $\hat D_1(\alpha)$ is
the unitary displacement operator.

From Eq.~\eqref{BSMatrix} together with Eq.~\eqref{reci} we easily obtain the
following relations:
\begin{eqnarray}
\hat{a}_{0}^{\dagger}=t'\hat{a}_{2}^{\dagger}+r'\hat{a}_{3}^{\dagger} \,\,
\mbox{and} \,\,
\hat{a}_{1}^{\dagger}=r\hat{a}_{2}^{\dagger}+t\hat{a}_{3}^{\dagger}.
\label{BS11}
\end{eqnarray}
Obviously two vacuum states at the input ports of the beam splitter
transform into vacuum states at the output ports:
$\ket{0}_0\ket{0}_1 \xrightarrow{BS} \ket{0}_2\ket{0}_3$. We use the
Baker-Campbell-Hausdorff formula \cite{Knight} and Eq. \eqref{BS11}
to see how the input state Eq.~\eqref{inputsingle} transforms into
the corresponding output state under the action of the beam
splitter:
\begin{align}
&\ket{1}_0 \ket{\alpha}_1 \xrightarrow{BS}  e^{\alpha(r \hat a_2^{\dagger} + t \hat a_3^{\dagger})-\alpha^*(r^* \hat a_2 + t^* \hat a_3)}(t' \hat a_2^{\dagger} + r' \hat a_3^{\dagger}) \ket{0}_2 \ket{0}_3  \notag \\ & = \hat D_2(r\alpha)\hat D_3(t\alpha)(t' \hat a_2^{\dagger} + r' \hat a_3^{\dagger}) \ket{0}_2 \ket{0}_3. \label{psiBS}
\end{align}

The density operator for the output state, Eq.~\eqref{psiBS}, reads
\begin{align}\label{}
\hat{\rho}_{23} &=  \hat D_2(r\alpha)\hat D_3(t\alpha) \hat{\tilde\rho}_{23} \hat D_3^{\dagger}(t\alpha) \hat D_2^{\dagger}(r\alpha),
\end{align}
where
\begin{equation}
\hat{\tilde\rho}_{23}=(t'\hat a_2^{\dagger} + r' \hat a_3^{\dagger})\ket{0}_2\ket{0}_3 \bra{0}_3\bra{0}_2 (t'^* \hat a_2 + r'^* \hat a_3).
\end{equation}
In fact $\hat{ \tilde\rho}_{23}$ would be the density operator of
the beam splitter output with vacuum instead of the coherent input
state. The two output modes (2 and 3) are in an entangled state. If
we consider only output mode 3 we have to find the reduced density
matrix by taking the partial trace over output 2, i.e.
\begin{align}\label{Trace}
\hat{\rho}_3 &= \Tr_2(\hat{\rho}_{23}) = \hat D_3(t\alpha)\Tr_2\left(\hat D_2(r\alpha)\hat{\tilde\rho}_{23} \hat D^{\dagger}_2(r\alpha)\right)\hat D^{\dagger}_3(t\alpha)\notag\\&= \hat D_3(t\alpha)\Tr_2\left(\hat D^{\dagger}_2(r\alpha)\hat D_2(r\alpha)\hat{\tilde\rho}_{23} \right)\hat D^{\dagger}_3(t\alpha) \notag\\&= \hat D_3(t\alpha) \sum_{n=0}^{\infty}\left<n\right|_2 \hat{\tilde\rho}_{23} \left| n \right>_2 \hat D^{\dagger}_3(t\alpha).
\end{align}
The operators $D_3(t \alpha)$ and $D_3^{\dagger}(t \alpha)$ are not
effected by the trace over mode 2, therefore we can put them outside
the trace. Further, in the second and third line of the last
equation we made use of the rule that the trace is invariant under
cyclic permutations, $\mbox{Tr}(ABC) = \mbox{Tr}(CAB)$, and the
unity relation $\hat D^{\dagger}(\alpha) \hat D(\alpha)=1$.
Inserting $\hat{\tilde\rho}_{23}$ into the last equation we get with
some basic boson algebra
\begin{equation}\label{densityBS}
\hat{\rho}_3= \abs{t'}^2 \ket{t\alpha}_3\bra{t\alpha}_3 +
\abs{r'}^2 \hat D_3(t\alpha) \ket{1}_3\bra{1}_3\hat D_3^{\dagger}( t\alpha).
\end{equation}
The result is a mixed state which is a convex combination of two
pure \emph{identically displaced} states: a coherent state
(displaced vacuum) and a displaced Fock state \cite{Oliv}, where the
displacement and the share of each pure state depends on the
reflectivity and the transmittivity of the beam splitter. For the
other output port we get the analogous result
\begin{equation}\label{densityBS2}
\hat{\rho}_2= \abs{r}^2 \ket{r\alpha}_2\bra{r\alpha}_2 +
\abs{t}^2 \hat D_2(r\alpha) \ket{1}_2\bra{1}_2\hat D_2^{\dagger}(r\alpha).
\end{equation}

\subsection{Wigner function}

With the simple form of the density operator for the output mode 3,
Eq.~\eqref{densityBS}, we can calculate its Wigner function. The
Wigner function is defined as
\cite{Wigner,WignerOptic,Schleich,Suda}
\begin{equation}\label{Wigner}
W_{\hat{\rho}}(q,p) = \frac{1}{2\pi \hbar} \int\limits_{-\infty}^{\infty}\matrixel{q-y/2}
{\hat{\rho}}{q+y/2}\mbox{exp}(iyp/ \hbar )dy,
\end{equation}
where $p$ and $q$ are the field quadratures. By inserting Eq.~\eqref{densityBS}
into Eq.~\eqref{Wigner} we see immediately that we get a sum of two Wigner
functions at the output port,
\begin{equation}\label{WignerBS}
W_{\hat{\rho}_3}  = \abs{t'}^2 W_{\hat \rho(\hat D(t\alpha)\ket{0})}
+ \abs{r'}^2 W_{\hat \rho (\hat D(t\alpha) \ket{1})},
\end{equation}
where $W_{\hat \rho(\hat D(t\alpha)\ket{0})}$ is the Wigner function
of the coherent state $\ket{t\alpha}$ and $W_{\hat \rho (\hat
D(t\alpha) \ket{1})}$ is the Wigner function of the displaced Fock
state $\hat D(t\alpha) \ket{1}$. These two individual functions are
well known \cite{Ulf} \cite{Oliv}. For the coherent state it is
\begin{equation}\label{Wigner_disp_Coh}
W_{\hat \rho(\hat D(t\alpha)\ket{0})}(q,p)=W_{\hat{\rho}(\ket{0})}(q',p')  =
\frac{1}{\pi \hbar} \, \mbox{exp}\left[-\left(\frac{q'}{q_0} \right)^2-\left(\frac{p'q_0}{\hbar}
\right)^2\right],
\end{equation}
and for the displaced Fock state it is
\begin{align}\label{Wigner_disp_Fock}
&W_{\hat \rho (\hat D(t\alpha) \ket{1})}(q,p) = W_{\hat{\rho}(\ket{1})}(q',p')  \notag \\ & =
-\frac{1}{\pi \hbar} \mbox{exp}\left [-\left(\frac{q'}{q_0} \right)^2-\left(\frac{p' q_0}{\hbar}
\right)^2\right ] \left[1+2\left(-\left(\frac{q'}{q_0}
\right)^2-\left(\frac{p' q_0}{\hbar}\right)^2\right)\right],
\end{align}
where $q_0 = \sqrt{\hbar/\omega}$. The displacement quadratures
\begin{align}\label{}
q' = q-\sqrt{2} q_0 (\operatorname{Re} \,t\alpha) \qquad \text{ and } \qquad
p' = p-\sqrt{2}\frac{\hbar}{q_0} (\operatorname{Im}\, t\alpha)
\end{align}
shift the minimum of the Fock state $W_{\hat{\rho}(\ket{1})}=-1$ (and the maximum of the coherent state $W_{\hat{\rho}(\ket{0})}=1$) to $q=\sqrt{2} q_0 (\operatorname{Re} \,t\alpha)$ and $p=\sqrt{2}\frac{\hbar}{q_0} (\operatorname{Im}\, t\alpha)$.

\begin{figure}[htbp] 
   \centering
   \subfigure[]
   {\includegraphics[width=0.49\textwidth]{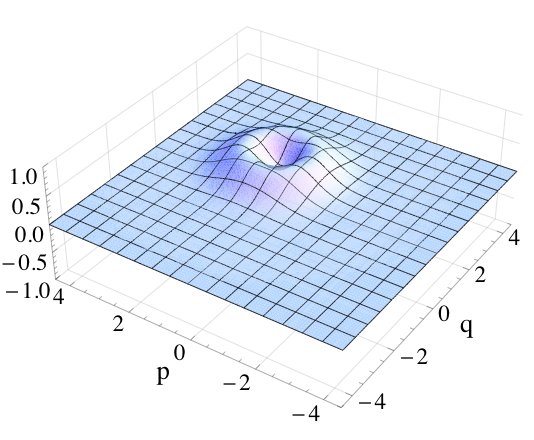}}
   \hfill
   \subfigure[]
   {\includegraphics[width=0.49\textwidth]{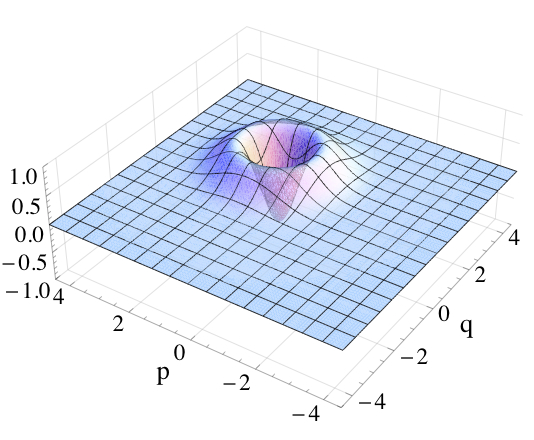}}
   \caption{Wigner function ($\times \pi$, $\hbar = \omega = 1$) of the beam splitter output 3, Eq.~\eqref{WignerBS}. It is a statistical mixture between a coherent state and a displaced Fock state.  (a) 50:50 beam splitter ($t = 1/\sqrt 2$) with the coherent input state $\ket{\alpha = \sqrt{2} e^{i \pi/4}}$.  (b) Highly reflective beam splitter ($t = 1/10$) with the coherent input state $\ket{\alpha = 10 e^{i \pi/4}}$. Note, that $t\alpha$ is identical for both (a) and (b). }
   \label{fig:Wigner}
\end{figure}

The Wigner function Eq.~\eqref{WignerBS} for the output field at
port 3 is depicted in Fig.~\ref{fig:Wigner} for two different cases.
In the first case we consider a 50:50 BS, where the Wigner function
is an equal mix between a coherent and a displaced Fock state.
The absolute minimum of this function is zero, situated at the center of the displaced Fock state, as can easily be seen from Eqs.~\eqref{WignerBS}-\eqref{Wigner_disp_Fock},
and
the total Wigner function is non-negative. In the second case we
consider a highly reflective (99:1) beam splitter comparable to the
experiment mentioned in the introduction \cite{Lvovsky}, in which a
displaced Fock state has been synthesized. However the high
reflectivity would lead to a smaller displacement at port 3, unless the
intensity of the incoming coherent state is increased to get an
identical factor $t\alpha$ for Fig.\ref{fig:Wigner}a and
Fig.\ref{fig:Wigner}b. Therefore the shift is identical, but due to
the different reflection and transmission coefficients of the BS the
relative weights of both input states are different. Indeed we see
from the Fig.\ref{fig:Wigner}b that the displaced Fock state is very
dominant with a pronounced minimum nearly reaching the original
value of -1.

We have thus established an exact theory for the experiment of
creating a displaced Fock state \cite{Lvovsky} instead of the theory
which the experiment was originally based on
\cite{Banaszek,Wallentowitz,Mancini} or the theory in \cite{Matteo}.

\subsection{Input: Coherent state and a general state}
For future theoretical considerations and experimental
demonstrations it might be useful to replace the ideal Fock state by
a general state
\begin{equation}\label{} \ket{\psi}_0 =
\sum_{m=0}^{\infty}c_m \ket{m}.
\end{equation}
The input state then reads
\begin{equation}\label{input}
\ket{\psi}_0 \ket{\alpha}_1 = \hat{D}_1(\alpha)
\sum_{m=0}^{\infty}c_m \frac{1}{\sqrt{m!}} \left(\hat
a_0^{\dagger}\right)^m  \ket{0}_0 \ket{0}_1\,\,\,.
\end{equation}
It is straightforward to see that the output mode 3 of the BS is
given by an equation analogous to Eq.~\eqref{Trace}, whereby the
density matrix $\hat{\tilde\rho}_{23}$ becomes
\begin{equation}\label{rho23}
\hat{\tilde\rho}_{23} = \sum_{m=0}^{\infty}\sum_{l=0}^{\infty}\frac{1}{\sqrt{m!l!}}c_mc_l^*(t'\hat a_2^{\dagger} + r' \hat a_3^{\dagger})^m\ket{0}_2\ket{0}_3 \bra{0}_3\bra{0}_2 (t'^* \hat a_2 + r'^* \hat a_3)^l.
\end{equation}
Similar to above $\hat{\tilde\rho}_{23}$ is the non displaced
density operator which would emerge at the output if the coherent
state at the input would be replaced by vacuum. Thus the coherent
input state displaces any state at the individual output port of a
beam splitter compared to a vacuum input (see Eq. \eqref{Trace}). In
the limit of a highly reflective beam splitter ($t \rightarrow 0$)
Eq. \eqref{rho23} becomes
$\hat{\tilde{\rho}}_{23}\rightarrow\ket{0}_2\ket{\psi}_3\bra{\psi}_3\bra{0}_2$.
If we furthermore consider a strong coherent state ($t \alpha$
finite), we see from Eq.~\eqref{Trace} that the effect on an
arbitrary input state is approximately a mere displacement by $t
\alpha$ of this state. A similar result has already been shown using
a different approach in \cite{Matteo} confirming our calculations.

\section{Interference at a Mach-Zehnder Interferometer}
\begin{figure}[htbp] 
   \centering
   {\includegraphics[width=0.60\textwidth]{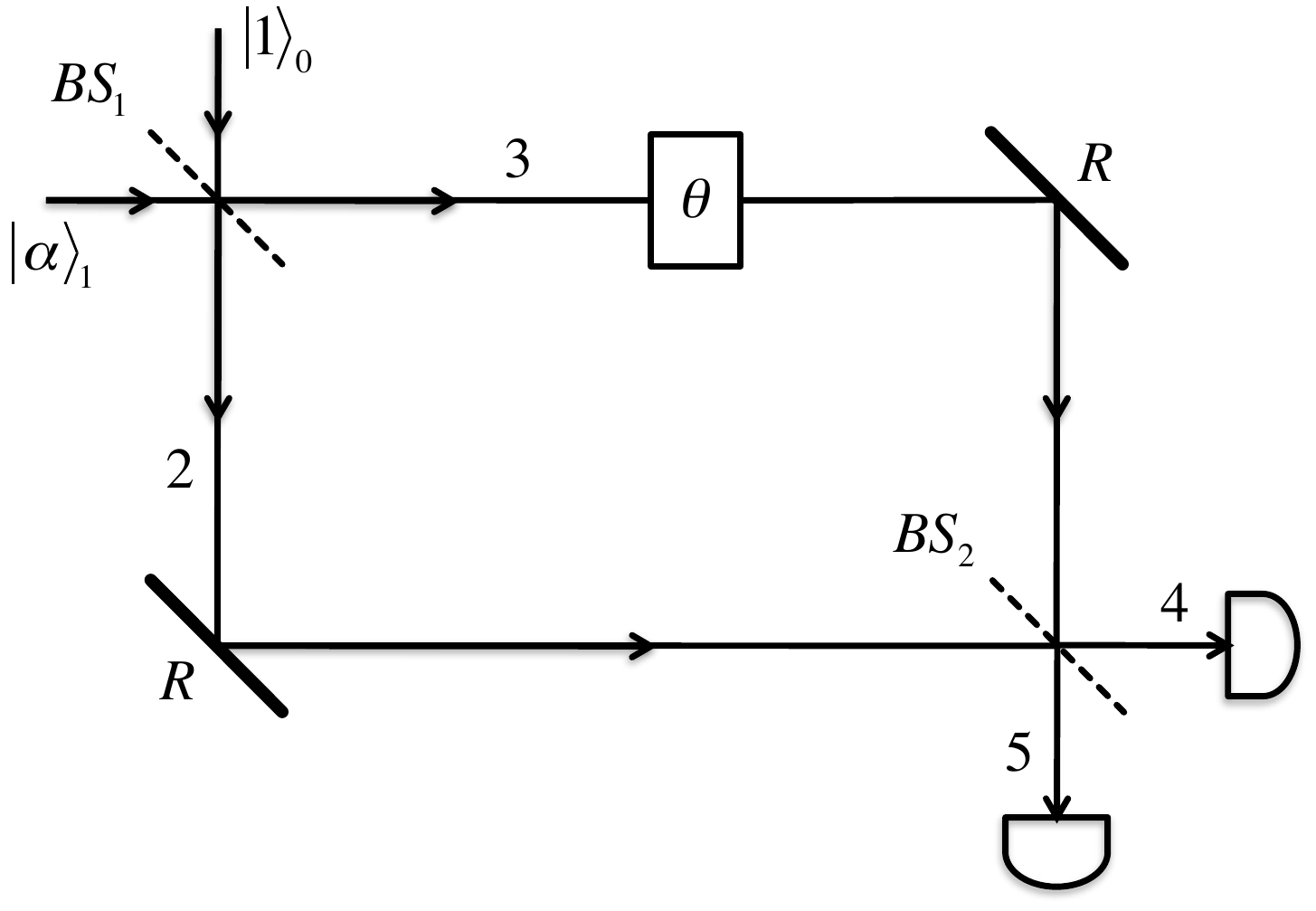}}
   \caption{Mach-Zehnder Interferometer (MZI) with a phase shift in path 3.
   The input states at $BS_1$ are a single photon Fock state $\ket{1}_0$
   and a coherent state $\ket{\alpha}_1$.}
   \label{fig:MZ}
\end{figure}

Now we consider a single-photon Fock state and a coherent state at
the input of a Mach-Zehnder interferometer (MZI) as shown in
Fig.~\ref{fig:MZ}. Again we want to calculate the quantum state at
the output ports.


\subsection{MZI scattering matrix}
As suggested in \cite{Yurke}, a MZI can be considered as a lossless
and passive four port device. The annihilation operators of the
field states transform similar to Eq.~\eqref{BSMatrix} and yield
\begin{equation}\label{MZquantumBS}
\begin{pmatrix}
\hat a_4\\
\hat a_5
\end{pmatrix} =
B_{MZI}
\begin{pmatrix}
\hat a_0\\
\hat a_1
\end{pmatrix}.
\end{equation}
The necessary and sufficient condition for the scattering matrix $B_{MZI}$ is that
it has to be unitary. In fact, since we have already considered a
general scattering matrix in the case of the beam splitter,
Eq.~\eqref{BSMatrix}, which essentially accounts for all passive and
lossless four port devices, the calculations from the last section
are generally valid. The scattering matrix of an arbitrary four port MZI, in particular, may be
represented as a composition of several unitary scattering matrices.


The scattering matrix for the MZI, depicted on Fig.\ref{fig:MZ} is composed of 3 unitary matrices,
two matrices corresponding to the two beam splitters (see Eq.~\eqref{BSMatrix}) and one matrix associated
with the phase shift, which in our case reads
\begin{equation}\label{}
U_{\theta}=
\begin{pmatrix}
1 &0\\
0 & e^{i\theta}
\end{pmatrix}.
\end{equation}
Accordingly the annihilation operators of the field states transform as
\begin{align}\label{MZItransfo}
\begin{pmatrix}
\hat{a}_4\\
\hat{a}_5
\end{pmatrix}
&= \begin{pmatrix}
t_2 &r_2'\\
r_2 &t_2'
\end{pmatrix}
\begin{pmatrix}
1 &0\\
0 & e^{i\theta}
\end{pmatrix}
\begin{pmatrix}
t_1' &r_1\\
r_1' &t_1
\end{pmatrix}
\begin{pmatrix}
\hat{a}_0\\
\hat{a}_1
\end{pmatrix}
\notag \\ &
= \begin{pmatrix}
t'_M & r_M \\
r'_M & t_M
\end{pmatrix}
\begin{pmatrix}
\hat{a}_0\\
\hat{a}_1
\end{pmatrix}
=B_{MZI}
\begin{pmatrix}
\hat{a}_0\\
\hat{a}_1
\end{pmatrix},
\end{align}
where in the last line we defined the following variables:
\begin{eqnarray}\label{MZIelements}
r_M = r_1 t_2+e^{i\theta}t_1 r_2', \nonumber\\
t_M = r_1 r_2+e^{i\theta}t_1 t_2', \nonumber\\
r'_M = t_1'r_2+e^{i\theta}r_1' t_2' , \nonumber\\
t'_M = t_1' t_2+e^{i\theta}r_1'r_2'.
\end{eqnarray}
The variables defined in Eq.~\eqref{MZIelements} obey the
reciprocity relations Eq.~\eqref{reci}, in particular
$|t_{M}|=|t'_{M}|$ and $|r_{M}|=|r'_{M}|$.

Here we add a short comment with respect to Fig.\ref{fig:MZ}. The
output ports $4$ and $5$ can serve as two paths of a second MZI
including a second phase shift, subsequently followed by a third MZI
of a similar type. Such a cascade of $3$ interferometers can operate
as preparation, distribution and measuring module of a simple linear
optical gate helpful in quantum computing systems. The resulting
scattering matrix of such a device would be a matrix product of $3$
matrices similar to $B_{MZI}$ used in Eq.~\eqref{MZItransfo}.

\subsection{MZI Output states}

The states at the output ports 4 and 5 of the MZI can now be calculated
analogously to the output states of the single beam splitter in the
previous section. Therefore according to Eq.~\eqref{psiBS} and by replacing the
beam splitter scattering matrix $B$ with the MZI scattering matrix
$B_{MZI}$ we get for the MZI output state
\begin{equation} \label{MZcoherentfockout1}
\ket{1}_0 \ket{\alpha}_1 \xrightarrow{MZI}  \hat D_4(r_M\alpha)\hat D_5(t_M\alpha)
(t'_M \hat a_4^{\dagger} + r'_M \hat a_5^{\dagger}) \ket{0}_4 \ket{0}_5.
\end{equation}
Further we get for the reduced density matrix at output port 5
(compare with Eq.~\eqref{densityBS}),
\begin{equation}\label{MZcoherentfockreduced}
\hat{\rho}_5 = \abs{t_M}^2 \ket{t_M \alpha}_5\bra{t_M \alpha}_5 +
\abs{r_M}^2 \hat D_5(t_M \alpha) \ket{1}_5\bra{1}_5\hat
D_5^{\dagger}( t_M \alpha)
\end{equation}
and output port 4
\begin{equation}\label{MZcoherentfockreduced4} \hat{\rho}_4 =
\abs{r_M}^2 \ket{r_M \alpha}_4\bra{r_M \alpha}_4 + \abs{t_M}^2 \hat
D_4(r_M \alpha) \ket{1}_4\bra{1}_4\hat D_4^{\dagger}( r_M \alpha).
\end{equation}

The result is again, as in the case of the beam splitter, a mixed
state between a coherent state (displaced vacuum) and a displaced Fock state, both displacements being identical. The
Wigner function of this state can be calculated using the analogue
of Eq.~\eqref{WignerBS}. The calculations where a coherent state and
a general state are incident on a MZI are equivalent to the beam
splitter case, so the output state is again merely displaced
compared to a MZI with vacuum input instead of the coherent input
state.

\subsection{Photon numbers}
Finally we calculate the average photon number at the output ports
of the MZI. For port 5 it is defined as
\begin{equation}\label{number}
\left< n \right>_5 = \Tr_5(\hat n_5 \hat \rho_5) = \sum_{n=0}^{\infty} \matrixel{n}{\hat a^{\dagger}_5 \hat a_5 \hat \rho_5}{n}_5.
\end{equation}
For the coherent state we obtain, of course,
\begin{equation}
 \sum_{n=0}^{\infty} \matrixel{n}{\hat a^{\dagger}_5  \hat a_5}{{t_M\alpha}}_5
 \Big\langle{t_M\alpha} \Big\vert n \Big\rangle_5 = \matrixel{{t_M\alpha}}
 {\hat a^{\dagger}_5 \hat a_5}{{t_M\alpha}}_5 = \abs{{t_M\alpha}}^2.
\end{equation}
For the displaced Fock state we obtain
\begin{align}
 &\sum_{n=0}^{\infty}\matrixel{n}{\hat a^{\dagger}_5 \hat a_5 D_5({t_M\alpha})}{1}_5
 \matrixel{1}{\hat D_5^{\dagger}({t_M\alpha})}{n}_5  \notag\\&=
 \matrixel{1}{\hat D_5^{\dagger}(t_M\alpha) \hat a^{\dagger}_5
 \hat a_5 \hat D_5(t_M\alpha)}{1}_5 =1+\abs{{t_M\alpha}}^2,
 \end{align}
where in the last step the commutation relation \cite{M&W} $\hat a
\hat D({t_M\alpha}) =$ $\hat D({t_M\alpha})(\hat a + {t_M\alpha}) $
has been used. With the last two equations and using that
$\abs{t_M}^2 + \abs{r_M}^2 = 1$ we can now calculate the average
photon number, Eq.~\eqref{number}, and get
\begin{equation}
\left< n \right>_5 = \abs{t_M}^2 \abs{{t_M\alpha}}^2 +
\abs{r_M}^2\left(1 + \abs{{t_M\alpha}}^2\right) = \abs{r_M}^2 +
\abs{{t_M\alpha}}^2.
\end{equation}
For port 4 we get analogously
\begin{equation}
\left< n \right>_4 = \abs{t_M}^2 + \abs{{r_M\alpha}}^2.
\end{equation}
Using this formalism the mean square deviation
$\langle(\Delta{}n)^{2}\rangle=\langle{}n^{2}\rangle-\langle{}n\rangle^{2}$
of photons can be calculated leading to
\begin{eqnarray}\label{MSDgeneral}
\langle(\Delta{}n)^{2}\rangle_{5}=|t_{M}\alpha|^{2}(1+2|r_{M}|^{2})+|r_{M}|^{2}|t_{M}|^{2}\,\,\,,
\nonumber\\
\langle(\Delta{}n)^{2}\rangle_{4}=|r_{M}\alpha|^{2}(1+2|t_{M}|^{2})+
|r_{M}|^{2}|t_{M}|^{2}\,\,\,
\end{eqnarray}
for the two output ports.

\subsection{Discussion of a MZI with balanced BSs}
\begin{figure}[htbp] 
   \centering
   \subfigure[]
   {\includegraphics[width=0.49\textwidth]{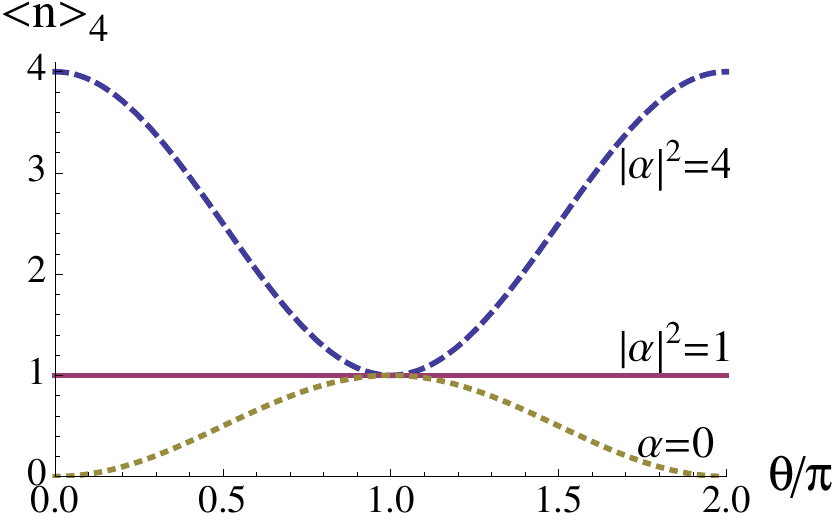}}
   \hfill
   \subfigure[]
   {\includegraphics[width=0.49\textwidth]{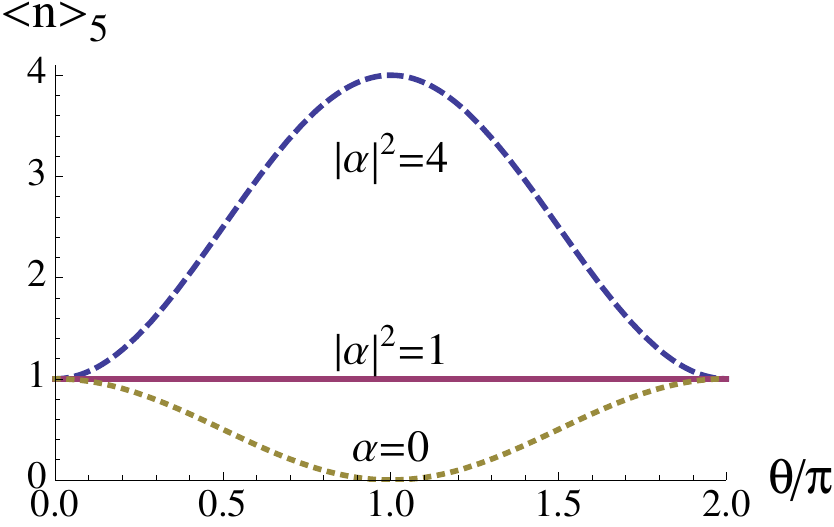}}
   \caption{Average photon number at output 4 (a) and output 5 (b) of the Mach Zehnder Interferometer with the coherent input states $\ket{\abs{\alpha} = 0}$, $\ket{\abs{\alpha} = 1}$ and $\ket{\abs{\alpha} = 2}$. Note that for $|\alpha|^{2}=1$ the mean
photon numbers are independent of $\theta$ and equal to 1.}
   \label{fig:photons}
\end{figure}

As a particular example, that can be realized easily in an experiment, we calculate the average photon numbers for the case
of two 50:50 dielectric layer beam splitters with a phase factor of
$e^{i\pi/2}=i$ for the reflected beams ($t_1=t'_1=t_2=t'_2 = 1/\sqrt
2, \, r_1 = r'_1=r_2=r_2' = i/\sqrt 2$) \cite{BSHamilton}.
In this case the transmittivity and the reflectivity are
$|t_{M}|^{2}=\sin^{2}(\theta/2)$ and
$|r_{M}|^{2}=\cos^{2}(\theta/2)$. Both are completely determined by
$\theta$.
The average photon numbers are then
\begin{equation}\label{n4}
\left< n \right>_4 = \sin^2(\theta/2) + \abs{\alpha}^2
\cos^2(\theta/2),
\end{equation}
\begin{equation}\label{n5}
\left< n \right>_5 = \cos^2(\theta/2)+ \abs{\alpha}^2
\sin^2(\theta/2)
\end{equation}
and the mean square deviations are given by
\begin{eqnarray}\label{MSD5050}
\langle(\Delta{}n)^{2}\rangle_{4}=
\frac{1}{4}\sin^{2}(\theta)+|\alpha|^{2}[\cos^{2}(\frac{\theta}{2})+\frac{1}{2}\sin^{2}(\theta)]
\,\,\,,\,\,\,
\langle(\Delta{}n)^{2}\rangle_{4}-\langle(\Delta{}n)^{2}\rangle_{5}=|\alpha|^{2}\cos(\theta)
\,.\,\,\,\,\,\,\,\,\,\,\,\,
\end{eqnarray}
Figure~\ref{fig:photons} shows the average photon numbers as a
function of the phase shift $\theta$ for different coherent input
states $\ket{\alpha}$. The maxima and minima in the graph indicate
the cases where the quantum states at the separate output ports are
either a pure coherent state or a pure Fock state.

The
technical advantage of the MZI (with two 50:50 BSs) over a single
beam splitter is, that ${t_M}$ and ${r_M}$ can simply be adjusted by
changing the phase $\theta$ in path 3. All
desired states could be generated in experiments without changing
the setup by selecting another beam splitter with the appropriate
transmittivity $\abs{t}^{2}$ and reflectivity $\abs{r}^{2}$.

Let us compare the result for $|\alpha|^{2}=1$ with the case when two separate single-photon Fock states are impinging on the input ports of an MZI. The first balanced $BS_{1}$ of the latter is reminiscent to the Hong-Ou-Mandel (HOM) effect \cite{HongOuMandel}.
After the $BS_{1}$ the state is
$\,i(|2\rangle_{2}|0\rangle_{3}+|0\rangle_{2}|2\rangle_{3})/\sqrt{2}\,$
which means that no coincidences appear behind $BS_{1}
$\,\cite{Knight}. Introducing a phase shift $\theta$ in path 3 and a
second $BS_{2}$ (see fig.~\ref{fig:MZ}) the wave function behind the
MZI reads
\begin{equation} \label{HOM}
|\psi\rangle=\frac{1}{2\sqrt{2}}\{(1-e^{2i\theta})[\,|0\rangle_{4}|2\rangle_{5}
-|2\rangle_{4}|0\rangle_{5}]+\sqrt{2}(1+e^{2i\theta})|1\rangle_{4}|1\rangle_{5}\}\,\,.
\end{equation}
The probability for coincidences is $W_{45}=\cos^{2}(\theta)$ while
the probabilities of measuring 2 photons at the output ports 4 and 5
are $W_{4}=W_{5}=\frac{1}{2}\sin^{2}(\theta)$. However, the mean
photon numbers $\langle{}n\rangle_{4}^{|\psi\rangle}$ and
$\langle{}n\rangle_{5}^{|\psi\rangle}$ are equal to 1 and therefore
independent of $\theta$. This can be compared with the results of the present paper (see Eqs.~\eqref{n4} and \eqref{n5}) in which the mean photon
numbers in output 4 and 5 also do not depend on $\theta$.
Anyway, one has to keep in mind that these two examples have completely different initial
conditions although the input mean photon numbers are the same in case of
$|\alpha|^{2}=1$. In fact, one has to compare the entangled quantum states of the HOM-effect Eq.~\eqref{HOM}
on the one hand with the MZI-state Eq.~\eqref{MZcoherentfockout1} on
the other hand and realize that these are completely different. By executing the trace operation for the separate
output ports, however,  entanglement is eliminated in both cases yielding the
identical outcomes for the mean photon numbers.


\section{Summary and outlook}
To sum up, we have derived simple analytical solutions for a quantum
state and its Wigner function at the output ports of a general
passive and lossless optical four port like a beam splitter, when a
coherent state and a single-photon Fock state are incident on the
input ports. These calculations could be of interest in the fields
of quantum cryptography based on continuous variables and optical
quantum computing. We have obtained a statistical mixture between of
a coherent state and a displaced Fock state and have derived the
corresponding Wigner functions. Furthermore, we have shown that a
coherent input state displaces the quantum state at the output port
of a beam splitter in phase space as compared with vacuum at the
input port, when in both cases a general state is incident on the
second input port. Additionally, we have analyzed the quantum states
behind a Mach-Zehnder interferometer and have evaluated mean photon
numbers at the output ports. It turns out that for an input state
$|1\rangle_{0}||\alpha|=1\rangle_{1}$ the mean photon numbers behind
the MZI do not depend on the phase shift $\theta$ inserted in the
interferometer.

In addition to the $3$-interferometer-cascade mentioned in section
III/A a further possible application of the presented formalism can
be proposed. Instead of the reflector $R$ in path $2$ of
Fig.~\ref{fig:MZ} a beam splitter can be inserted in such a way that
beam $2$ is both reflected and transmitted. The beam splitter
provides also an additional input. A second MZI then can be added
below the first one. In particular these two interferometers have
one beam path in common generating a so-called two-loop
interferometer which consists of $4$ beam splitters and $2$
reflectors providing $3$ input and $3$ output ports. Such a device
enables the superposition of three wave functions and will be
theoretically investigated in a next step.

The results of this paper can serve as a basis for further phase-space
investigations of higher Fock states in combination with coherent,
squeezed or thermal states as input states used in current optical
setups. Those setups (BS, MZI, Phase gate, CNOT gate) are building
blocks for linear optical quantum computers.

\bibliographystyle{utphys}

\bibliography{literatur}

\providecommand{\href}[2]{#2}\begingroup\raggedright\begin{thebibliography}{10}

\bibitem{quantumcryptography}
M.~D\v{u}sek, N.~L\"utkenhaus, and M.~Hendrych, ``Quantum Cryptography,'' {\em
  Progress in Optics} {\bfseries 49} (2006) 381,
  \href{http://arxiv.org/abs/quant-ph/0601207}{{\ttfamily quant-ph/0601207}}.

\bibitem{continuousvariableQKD}
J.~Lodewyck, M.~Bloch, and R.~G.-P. et~al., ``Quantum key distribution over 25
  km with an all-fiber continuous-variable system,'' {\em Phys. Rev. A}
  {\bfseries 76} (2007) 042305,
  \href{http://arxiv.org/abs/0706.4255}{{\ttfamily arXiv:0706.4255
  [quant-ph]}}.

\bibitem{Chuang}
M.~A. Nielsen and I.~L. Chuang, {\em Quantum Computation and Quantum
  Information}.
\newblock Cambridge University Press, Cambridge, 2000.

\bibitem{Gisin}
N.~Gisin, G.~Ribordy, W.~Tittel, and H.~Zbinden, ``Quantum Cryptography,'' {\em
  Rev. Mod. Phys} {\bfseries 74} (2002) 145,
  \href{http://arxiv.org/abs/quant-ph/0101098}{{\ttfamily quant-ph/0101098}}.

\bibitem{quantcomputing1}
J.~L. O'Brien, ``Optical Quantum Computing,'' {\em Science} {\bfseries 318}
  (2007) 1567, \href{http://arxiv.org/abs/0803.1554}{{\ttfamily arXiv:0803.1554
  [quant-ph]}}.

\bibitem{quantcomputing2}
A.~Politi, M.~J. Cryan, J.~G. Rarity, S.~Yu, and J.~L. O'Brien,
  ``Silica-on-Silicon Waveguide Quantum Circuits,'' {\em Science} {\bfseries
  320} (2008) 646, \href{http://arxiv.org/abs/0802.0136}{{\ttfamily
  arXiv:0802.0136 [quant-ph]}}.

\bibitem{M&W}
L.~Mandel and E.~Wolf, {\em Optical Coherence and Quantum Optics}.
\newblock Cambridge Univ. Press, Cambridge, 1995.

\bibitem{BS2}
U.~Leonhardt, ``Quantum Physics of Simple Optical Instruments,'' {\em
  Rept.Prog.Phys.} {\bfseries 66} (2003) 1207,
  \href{http://arxiv.org/abs/quant-ph/0305007}{{\ttfamily quant-ph/0305007}}.

\bibitem{Yurke}
B.~Yurke, S.~L. McCall, and J.~R. Klauder, ``{SU}(2) and {SU}(1)
  interferometers,'' {\em Phys. Rev. A} {\bfseries 33} (1986) 4033.

\bibitem{Knight}
C.~C. Gerry and P.~L. Knight, {\em Introductory Quantum Optics}.
\newblock Cambridge Univ. Press, Cambridge, 2005.

\bibitem{HongOuMandel}
C.~K. Hong, Z.~Y. Ou, and L.~Mandel, ``Measurement of Subpicosecond Time
  Intervals between Two Photons by Interference,'' {\em Phys. Rev. Lett.}
  {\bfseries 59} (1987) 2044.

\bibitem{HOM88}
C.~K. Hong, Z.~Y. Ou, and L.~Mandel, ``Interference between a fluorescent
  photon and a classical field: An example of nonclassical interference,'' {\em
  Phys. Rev. A} {\bfseries 37} (1988) 3006.

\bibitem{Bali93}
S.~Bali, F.~A. Narducci, and L.~Mandel, ``Coherence and elastic scattering in
  resonance fluorescence,'' {\em Phys. Rev. A} {\bfseries 47} (1993) 5056.

\bibitem{Lvovsky}
A.~I. Lvovsky and S.~Babichev, ``Synthesis and Tomographic Characterization of
  the displaced Fock state of light,'' {\em Phys. Rev. A} {\bfseries 66} (2002)
  011801, \href{http://arxiv.org/abs/quant-ph/0202163}{{\ttfamily
  quant-ph/0202163}}.

\bibitem{BSHamilton}
M.~W. Hamilton, ``Phase shifts in multilayer dielectric beam splitters,'' {\em
  Am. J. Phys.} {\bfseries 68} (2000) 186.

\bibitem{Stokes}
G.~G. Stokes {\em Cambridge \& Dublin Math. J.} {\bfseries 4} (1849) 1.

\bibitem{Oliv}
F.~A.~M. de~Oliveira, M.~S. Kim, P.~L. Knight, and V.~Bu\v{s}ek, ``Properties
  of displaced number states,'' {\em Phys. Rev. A} {\bfseries 41} (1990) 2645.

\bibitem{Wigner}
E.~Wigner, ``On the Quantum Correction For Thermodynamic Equilibrium,'' {\em
  Phys. Rev.} {\bfseries 40} (1932) 749.

\bibitem{WignerOptic}
M.~Hillary, R.~F. O'Connell, M.~O. Scully, and E.~P. Wigner, ``Distribution
  Function in Physics: Fundamentals,'' {\em Phys. Rep} {\bfseries 106} (1984)
  121--167.

\bibitem{Schleich}
W.~P. Schleich, {\em Quantum Optics in Phase Space}.
\newblock Wiley-Vch, Berlin, 2001.

\bibitem{Suda}
M.~Suda, {\em Quantum Interferometry in Phase Space}.
\newblock Springer, Berlin, 2006.

\bibitem{Ulf}
U.~Leonhardt, {\em Measuring the Quantum State of Light}.
\newblock Cambridge Univ. Press, Cambridge, 1997.

\bibitem{Banaszek}
K.~Banaszek and K.~Wodkiewicz, ``Direct Probing of Quantum Phase Space by
  Photon Counting,'' {\em Phys. Rev. Lett.} {\bfseries 76} (1996) 4344,
  \href{http://arxiv.org/abs/atom-ph/9603003}{{\ttfamily atom-ph/9603003}}.

\bibitem{Wallentowitz}
S.~Wallentowitz and W.Vogel, ``Unbalanced Homodyning for Quantum State
  Measurements,'' {\em Phys. Rev. A} {\bfseries 53} (1996) 4528.

\bibitem{Mancini}
S.~Mancini, P.~Tombesi, and V.~Manko, ``Density Matrix from Photon Number
  Tomography,'' {\em Europhys. Lett.} {\bfseries 37} (1997) 79,
  \href{http://arxiv.org/abs/quant-ph/9612004}{{\ttfamily quant-ph/9612004}}.

\bibitem{Matteo}
M.~G.~A. Paris, ``Displacement operator by beam splitter,'' {\em Phys. Lett. A}
  {\bfseries 217} (1996) 78--80.

\end{thebibliography}\endgroup

\end{document}